\documentstyle[emulateapj]{article}
\begin{document}

\def\lsim{\mathrel{\hbox{\rlap{\hbox{\lower4pt\hbox{$\sim$}}}\hbox{$<$}}}}
\def\gsim{\mathrel{\hbox{\rlap{\hbox{\lower4pt\hbox{$\sim$}}}\hbox{$>$}}}}

\title{Halo Substructure and Disk Heating in a $\Lambda$CDM universe}
\author{Andreea S. Font$^1$, Julio F. Navarro$^{1,}$\altaffilmark{3}, Joachim
Stadel$^{1,}$\altaffilmark{4}, and  Thomas Quinn$^2$}

\affil{$^1$Department of Physics and Astronomy, University of Victoria, 
 Victoria, BC, V8P 1A1, Canada}

\affil{$^2$Department of Astronomy, University of Washington, Seattle, WA
98195, USA.}

\altaffiltext{3}{CIAR Scholar and Alfred P. Sloan Research Fellow}

\altaffiltext{4}{CITA National Fellow and PIMS Fellow}

\begin{abstract}
We examine recent suggestions that substructure in cold dark matter (CDM) halos
may be in conflict with the presence of thin, dynamically fragile stellar
disks. N-body simulations of an isolated disk/bulge/halo model of the Milky Way
that includes several hundred dark matter satellites with masses, densities and
orbits derived from high-resolution cosmological CDM simulations indicate that
substructure at $z=0$ plays only a minor dynamical role in the heating of the
disk over several Gyrs. This is because the orbits of satellites in present-day
CDM halos seldom take them near the disk, where their tidal effects are
greatest. Unless the effects of substructure are very different at earlier
times, our models suggest that substructure might not preclude virialized CDM
halos from being acceptable hosts of thin stellar disks like that of the Milky
Way.
\end{abstract}

\keywords{N-body simulations; galaxies: evolution, structure; cosmology: dark matter}

\section{Introduction}

Over the past few years cosmological N-body simulations have achieved
unprecedented resolution. One recent highlight of these studies has been the
discovery that during the merger events that characterize the assembly and
growth of dark matter halos the central regions of accreted halos may survive
for several orbital times as self-bound entities in the parent halo (Klypin et
al 1999, Moore et al 1999, Ghigna et al 2000).

This population of surviving halo cores, or ``subhalos'', typically contributes
less than $\sim 10\%$ of the total mass of the system, with the bulk of the mass
in a smooth monolithic structure, as envisioned by White \& Rees (1978). Despite
the small fraction of the total mass they make up, at any given time a large
number of subhalos are expected within the virialized region of a cold dark
matter halo. In a galaxy cluster, it is tempting to identify these subhalos as
dynamical tracers of the individual galaxy population (Moore et al 1999,
hereafter M99), but in the case of a galaxy-sized halo such association would
lead to the expectation of several hundred satellites in the vicinity of the
Milky Way. For example, M99 find that up to $500$ satellites with circular
velocities exceeding $\sim 10$ km/s may have survived within $\sim 300$ kpc from
the center of the Galaxy.  Comparing this large number with the dozen or so
known Milky Way satellites with comparable velocity dispersion clearly implies
that most subhalos must have failed to form a significant number of stars if
these models are to match observations.

The lack of a significant luminous component associated with most subhalos was
suggested several years ago by semianalytic models of galaxy formation (see,
e.g., Kauffmann, White \& Guiderdoni 1993), and is analogous to the need to
prevent a large number of dwarf galaxies from forming in low-mass halos first
noted by White \& Rees (1978) and confirmed later by semianalytic models of
galaxy formation (White \& Frenk 1991, Cole et al 1994, Somerville \& Primack
1999). Energetic mechanisms that operate more efficiently on low-mass systems,
such as feedback from evolving stars (White \& Rees 1978), as well as heating by
an ionizing UV background (Efstathiou 1992, Bullock, Kravtsov \& Weinberg 2000),
are usually invoked to explain this decoupling of luminous and dark components
on small mass scales and are a staple of hierarchical galaxy formation models.

Even if luminous galaxies fail to ``turn on'' in most subhalos a potential
difficulty with the substructure expected in CDM halos has been cited by M99:
the fluctuating gravitational potential induced by the clumpy structure of the
halo may act to heat and thicken fragile stellar disks beyond observational
constraints (see, e.g., Toth \& Ostriker 1992). M99 estimate that the energy
pumped into a disk like that of the Milky Way by subhalos may add up to a
significant fraction of the binding energy of the disk over a Hubble
time. Estimating analytically the heating rate due to the cumulative effect of
disk/subhalo collisions is, however, quite uncertain, and in practice simple
calculations that neglect the self-gravitating response of the disk often
overestimate the effects of heating by infalling satellites (Huang \& Carlberg
1997, Vel{\'{a}}zquez \& White 1999).

In this {\it Letter} we present the results of an attempt to gauge the dynamical
effects of substructure on the dynamical evolution of thin stellar disks
embedded in dark matter halos. Our approach uses the mass spectrum, orbit
distribution, and internal structure of subhalos identified in cosmological
simulations of galaxy-sized CDM halos.

\section{Substructure in CDM halos}

The mass function of substructure halos, their internal structure, and the
parameters of their orbits are the main properties of the subhalo population
that determine the tidal effects of substructure on stellar disks.  Subhalos
discussed here and in M99 have been identified using the groupfinder
SKID{\footnote{{\tt
http://www-hpcc.astro.washington.edu/tools/skid.html}}}. SKID's main input
parameter is a characteristic length scale, $l_S$, which determines the
characteristic size of the smallest groups and, indirectly, the minimum
separation between clumps.


{\epsscale{1.}
\plotone{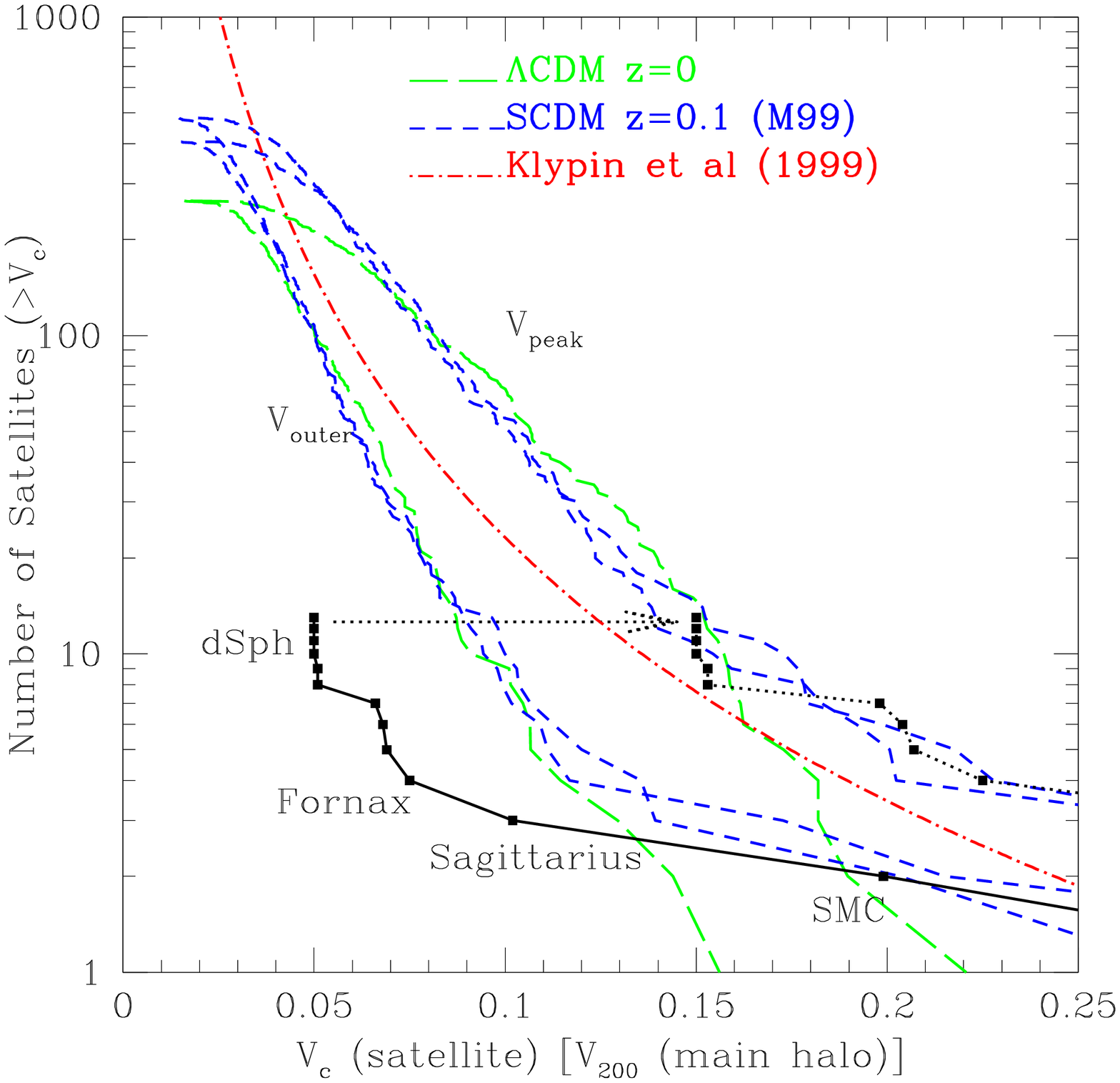}}
{\small {\sc Fig.}~1.
Cumulative circular velocity function of substructure halos.  Velocities are
normalized to the circular velocity of the parent halo measured at the virial
radius. Only subhalos within the virial radius of the main halo are
included. Short-dashed curves correspond to two galaxy-sized halos formed in a
SCDM universe, derived from data kindly provided by M99. Long-dashed curves
correspond to a halo formed in the $\Lambda$CDM cosmogony. The two sets of
curves, labeled $V_{\rm peak}$ and $V_{\rm outer}$, indicate the result of
adopting the maximum circular speed or the circular velocity at the outermost
bound radius, respectively. The dot-dashed curve illustrates the results
reported by Klypin et al (1999). Filled squares correspond to Milky Way
satellites assuming that stars in dwarf spheroidals are on isotropic orbits in
isothermal potentials. The dotted line corresponds to the same data but where
circular velocities have been increased by a factor of three to reflect
uncertainties in the structure of dark halos surrounding dwarf spheroidals (see,
e.g., White 2000).}\bigskip

The short-dashed curves in Figure 1 show the subhalo velocity function
corresponding to two galaxy-sized dark matter halos formed in the ``Local
Group'' run reported by M99. The ``Local Group'' simulation assumed the former
``standard'' CDM cosmogony (SCDM: $\Omega_M = 1,
\Omega_{\Lambda}= 0, h = 0.5, \sigma_8=0.7$). The two halos shown here have
$V_{200} \sim 193$ and $172$ km s$^{-1}$, respectively, and each has roughly
$10^6$ particles within $r_{200}$, the ``virial'' radius within which the mean
density is $200$ times the critical density for closure.

Circular velocity (instead of mass) is used to characterize subhalos because of
its weaker dependence on the exact way in which substructure is
identified. Still, subhalo circular velocities change with
radius, and there is no unique way of defining subhalo circular velocities. SKID
reports $V_{\rm peak}$ and $V_{\rm outer}$, which correspond to the maximum
circular speed within the subhalo, and to its value at the outermost bound
radius, respectively. These two measures may differ in some cases by up to $\sim
50\%$, implying rather different substructure velocity functions, as can be seen
in Figure 1.  The substructure function presented in M99's Figure 2 is
consistent with the set of curves labeled $V_{\rm peak}$ in Figure
1{\footnote{Note that the caption to M99's Figure 2 suggests, apparently in
error, that $V_{\rm outer}$ is used there rather than $V_{\rm peak}$.}}.

{\epsscale{1.0}
\plotone{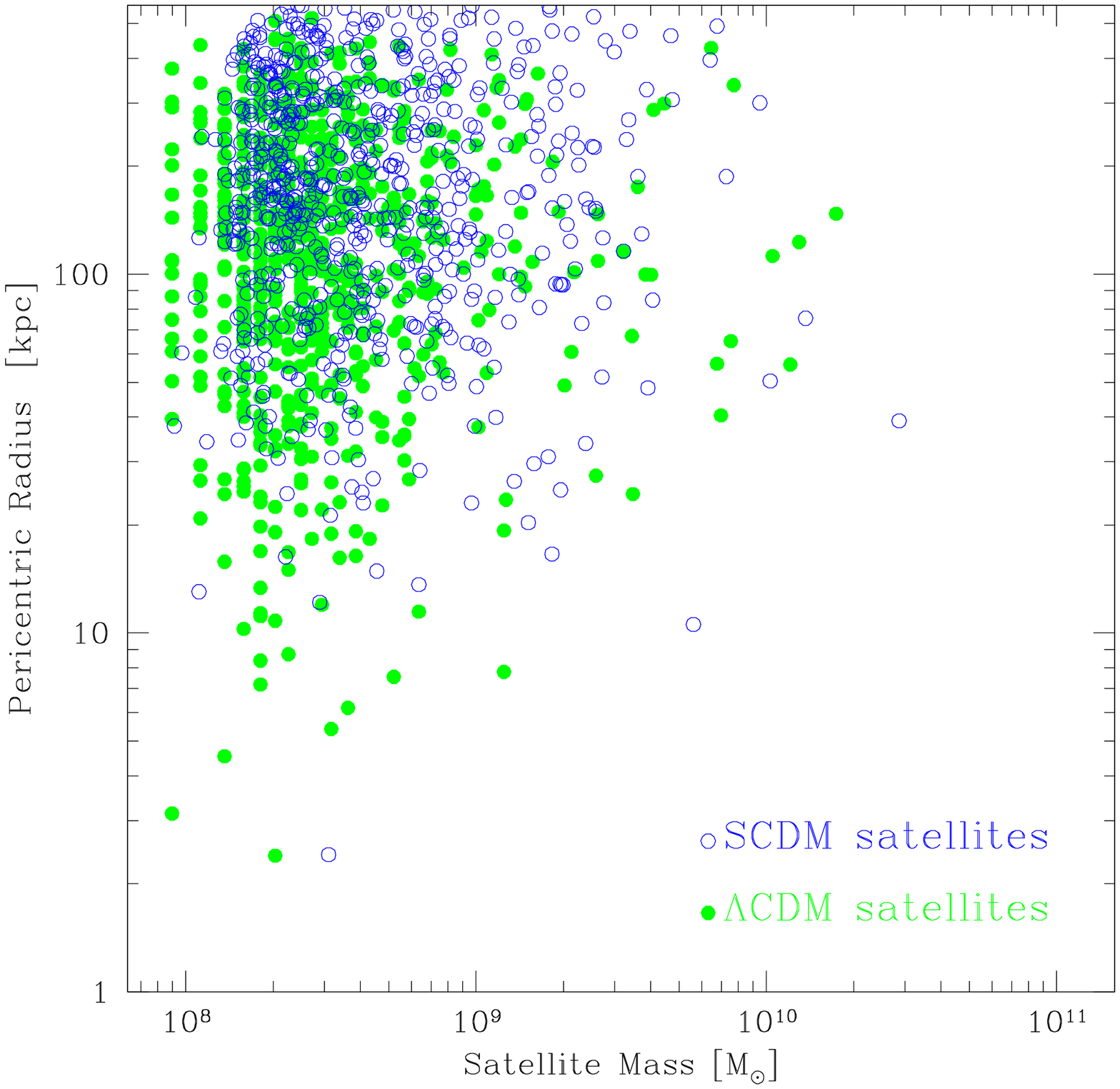}}
{\small {\sc Fig.}~2. Mass vs pericentric radii for subhalos identified in two CDM halos, after scaling to the virial radius and circular velocity of our Milky Way disk model.} 
\bigskip

As pointed out by M99 and Klypin et al (1999), the substructure velocity
function, scaled to the circular speed of the parent halo, is roughly
independent of the mass of the parent halo. The long dashed curves in Figure 1
indicate that this function is also rather insensitive to the cosmological model
adopted. These curves correspond to a $\sim 205$ km s$^{-1}$ halo formed in the
$\Lambda$CDM cosmogony ($\Omega_M = 0.3, \Omega_{\Lambda} = 0.7, h = 0.65,
\sigma_8=0.9$). This system was selected at random from a large simulation of a
$32.5 \, h^{-1}$ Mpc periodic box, and resimulated at high resolution using
PKDGRAV. The procedure for resimulation is analogous to that described in Eke,
Navarro \& Steinmetz (2001), and is similar (but not identical) to that used for
the SCDM ``Local Group'' halos.  The mass and force resolution are similar in
the SCDM and $\Lambda$CDM runs, and they have both been carried out with the
same N-body code (PKDGRAV).

The insensitivity to cosmogony of the substructure velocity function was first
reported by Klypin et al (1999). Their results are shown by the dot-dashed line
in Figure 1 and are in reasonable agreement with our determination, provided
that the circular velocities computed by their substructure-finding algorithm
fall, as it seems plausible, between $V_{\rm peak}$ and $V_{\rm outer}$.

How does the subhalo velocity function compare with the observed number of
satellites in the vicinity of the Milky Way? The solid squares in Figure 1
(joined by a solid line) illustrate the cumulative number of known Milky Way
satellites as a function of the circular velocity of their halos, as plotted by
M99. Here, circular velocities are derived for the halos of dwarf spheroidals
assuming that stars in these systems are on isotropic orbits in isothermal
potentials. This is a plausible, but nevertheless questionable, assumption. Dark
halos differ significantly from simple isothermal potentials, and numerical
simulations indicate that circular velocities decrease substantially near the
center. If stars populate the innermost regions of subhalos their velocity
dispersions may substantially underestimate the subhalo peak circular
velocities. 

This has been discussed by White (2000), who finds, using the mass model
proposed by Navarro, Frenk \& White (1996, 1997, NFW), that dwarf spheroidals
may plausibly inhabit potential wells with circular velocities up to a factor of
$3$ times larger than inferred under the isothermal assumption.  Such correction
may reconcile, at the high mass end, the Milky Way satellite velocity function
with the subhalo function, as shown by the dotted line in Figure 1.
Thus, it is possible that the number of {\it massive} satellites expected in the
CDM scenario might not be in gross conflict with observation.  

On the other hand, this does not erase the large number of low mass ``dark''
satellites that should inhabit the main halo. Can thin stellar disks survive
unscathed in this clumpy environment? The tidal heating rate by substructure can
be shown to scale as $dE/dt \propto \int n(m_{s}) \, m_{s}^2 \, dm_{s}$, where
$m_s$ is the subhalo mass (White 2000). Since, according to numerical
simulations, $n(m_s)\propto m_s^{-1.8}$ (Ghigna et al 2000, Springel et al
2000), the tidal effects of substructure are dominated by the few most massive
subhalos. Stochastic effects are therefore expected to dominate the heating rate
and a statistically significant sample of systems should be simulated in order
to obtain conclusive results.  The results reported here (which are based on two
realizations) should then be viewed as preliminary.

\section{The Evolution of Stellar Disks in a $\Lambda$CDM halo}
\subsection{ The Galaxy Model}

We have carried out a series of numerical simulations that follow the evolution
of a disk of particles within a dark matter halo with substructure similar to
that of the high-resolution $\Lambda$CDM halo alluded to in Figure 1. Since we
are interested in isolating the effects of substructure on the disk, we choose
to minimize the influence of other effects (such as the triaxiality of the halo,
spiral structure, disk warping) by carefully constructing an equilibrium
disk/bulge/halo model following the procedure outlined by Hernquist (1993). We
choose parameters so that the model reproduces many of the observational
characteristics of the Milky Way galaxy, as in Vel{\'{a}}zquez \& White
(1999). The disk is assumed to be exponential, with a total mass of $5.6 \times
10^{10} M_{\odot}$, exponential scale length of $3.5$ kpc, and vertical scale
height of $700$ pc. The bulge is modeled as a Hernquist (1993) model with total
mass $1.87 \times 10^{10} M_{\odot}$ and $525$ pc scale radius. Finally, we
adopt a non-singular isothermal model for the dark halo, with a total mass of
$3.09 \times 10^{12} M_{\odot}$, a core radius $r_c=3.5$ kpc, and exponentially
truncated at a distance of $300$ kpc from the center. The scaled circular
velocity profile of the galaxy model closely resembles that of the $\Lambda$CDM
halo out to $\sim 600$ kpc, about twice its virial radius.

The galaxy model is first evolved without substructure halos in an attempt to
assess the deviations from equilibrium induced by noise in the particle
distribution. N-body disks are notoriously unstable, and substantial numbers of
particles are needed to ensure stability over several rotation periods. The
simulations we report here use $40,000$ particles in the disk, $13,333$ in the
bulge, and $2.2 \times 10^6$ in the halo, all of equal mass, $m_p=1.4 \times
10^{6} M_{\odot}$. All simulations were run with PKDGRAV using a spline
softening length of $175$ pc and a hierarchy of timesteps, the smallest of which
is typically less than $10^5$ yrs.

\subsection{Including Substructure}

Substructure halos, despite their large numbers, make up a small fraction of the
total mass of the system, implying that it is possible to add subhalos to our
galaxy model without seriously compromising the global equilibrium of the
system. We include subhalos using the following procedure. We first run SKID on
the CDM halos discussed in \S2, using the characteristic linking-length scales
that maximize the number of subhalos found within the virial radius. This yields
$685$ and $737$ satellites with masses exceeding $\sim 10^8 M_{\odot}$ within $2
\, r_{200}$ for the $\Lambda$CDM and SCDM halo, respectively.

We choose the peak circular velocity, $V_{\rm peak}$, and the radius at which it
occurs, $r_{\rm peak}$, to characterize each subhalo. The values of $r_{\rm
peak}$ and $V_{\rm peak}$ are then used to generate, for each of the satellites,
an equilibrium NFW model with particle masses equal to those used in the
disk/halo galaxy model\footnote{Note that these two parameters fully specify an
NFW halo model.}. In order to ensure a finite mass we truncate each NFW model at
$5 \, r_{\rm peak}$. The positions and velocities of each satellite are scaled
so that they correspond to the same fractions of $r_{200}$ and $V_{200}$,
respectively, in both the parent CDM halos and in the galaxy model. This
procedure ensures that the subhalos have approximately the same masses,
densities and orbits as those in the CDM halos. Figure 2 shows the masses and
pericentric distances of the subhalos within $2 \, r_{200}$ included in the
simulations. We include satellites within twice the virial radius because many
of them have time to reach the pericenter of their orbits (which lie within
$r_{200}$) during the timespan of our simulations. Note that there are no
satellites with masses exceeding $10^{11} M_{\odot}$ and that only a few
satellites are on orbits that take them within the solar circle ($8.5$ kpc). All
of those satellites have masses below $10^9 M_{\odot}$. Figure 3 shows an
edge-on view of the disk/bulge/halo/subhalo model.

\subsection{Disk Heating}

We evolve the disk model first for about $3.5$ Gyrs without including
substructure in order to quantify the heating rate due to noise in the particle
distribution.
The ($R$, $z$, $\phi$) velocity dispersions of disk particles at the solar
circle (i.e., $8.5$ kpc from the center) grow from ($31$, $27$, $26$) km
s$^{-1}$ to ($43$, $30$, $33$) km s$^{-1}$ over the same period (see solid lines
in Figure 4). Quantifying this heating rate by the usual expression
$\sigma_{tot}^2=\sigma_0^2+Dt$, we find $D\sim 200$ km$^2$ s$^{-2}$ Gyr$^{-1}$, about
a factor of two less than inferred for stars in the solar neighborhood from the
age-velocity dispersion (Wielen 1977, Lacey \& Ostriker 1985, Fuchs et al
2000). 

The two-body heating rate in our equilibrium stellar disk model is thus low
compared with the actual heating experienced by stars in the disk of the
Galaxy. This implies that our disk model is stable enough to verify numerically
whether substructure in the halo leads to heating rates inconsistent with
observational constraints. This is shown by the dashed lines in Figure 4, which
show the evolution of the disk velocity dispersion when the substructure halos
are added to the system. Clearly, the heating rate is approximately the same
with and without substructure, a result that may be traced to the fact that
there are no satellites more massive than $10^{11} \, M_{\odot}$ and that their
orbits seldom take them near the disk. If substructure in the halos we consider
is representative of galaxy-sized CDM halos (and we have no reason to suspect it
is not), this would imply that tidal heating rates of thin stellar disks by
substructure halos may be consistent with the observational evidence.

{\epsscale{1.}
\plotone{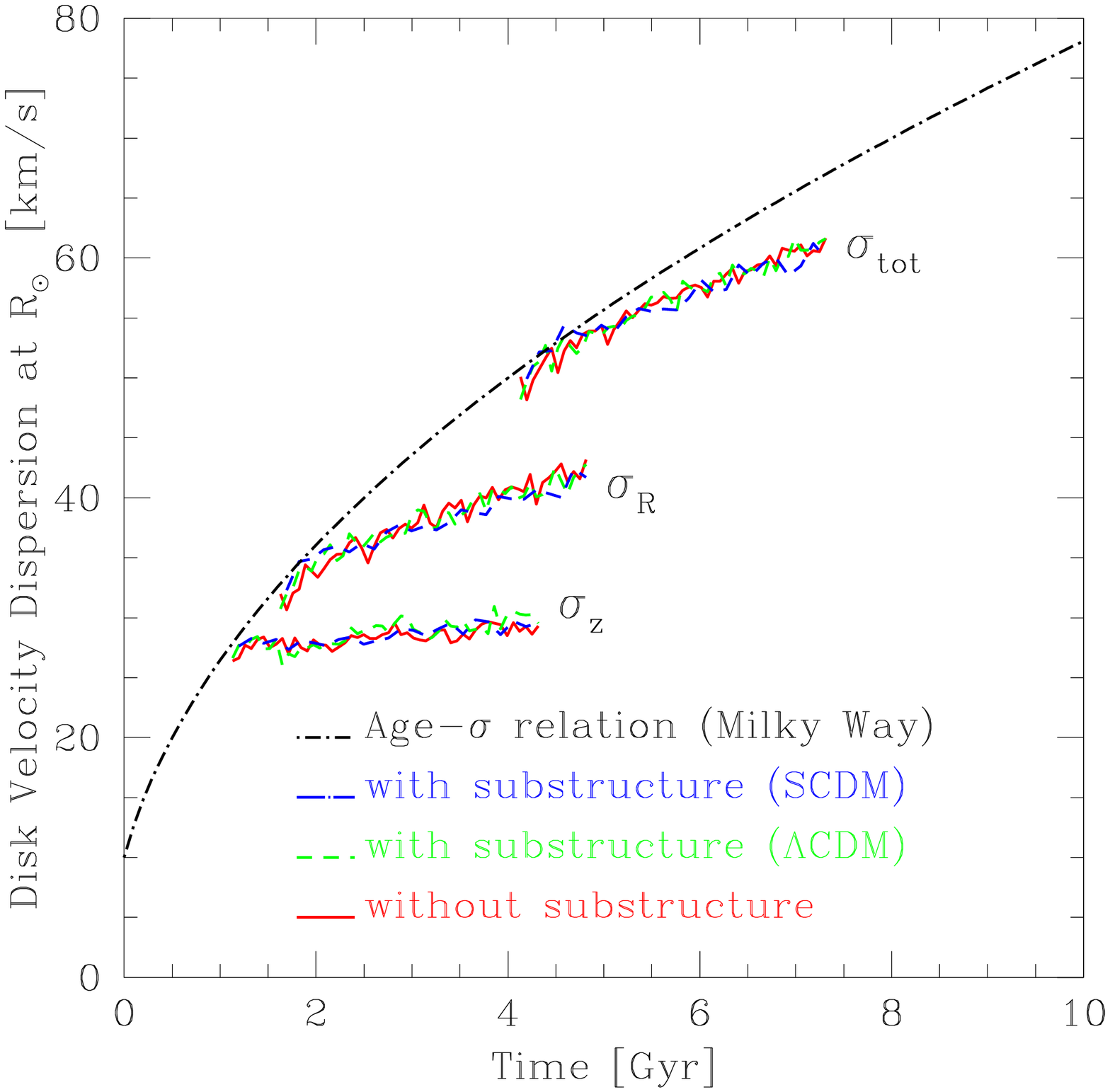}}
{\small {\sc Fig.}~4. Stellar disk velocity dispersion as a function of time 
in our Milky Way model compared with the age-velocity dispersion relation 
in the solar neighborhood, as compiled recently by Fuchs et al (2000, dot-dashed curve). Solid (dashed) lines indicate the heating rate of our disk/bulge/halo model evolved for about $3.5$ Gyrs excluding (including)
substructure. ``Starting'' times for simulation data are arbitrary and have been chosen to lie on the age-$\sigma_{tot}$ relation for the solar
neighborhood. }
\bigskip

\section{Discussion and Conclusions}

Our results suggest that concerns regarding excessive tidal heating of thin
stellar disks by substructure in CDM halos may be less serious than previously
thought. Once subhalos with the mass, structure, and orbital distribution
expected in a CDM universe at $z \sim 0$ are considered, direct numerical
simulation shows that thin stellar disks at the center of such halos may survive
undisturbed by collisions with subhalos. This is because subhalo masses in
galaxy-sized halos are typically below $10^{10}$-$10^{11} M_{\odot}$ and have
orbits that usually do not take them close to the disk. We conclude that the
substructure observed in virialized CDM halos is not clearly inconsistent with
the existence of thin stellar disks such as that of the Milky Way.

These conclusions are subject to a number of caveats. The most obvious one is
that our study explores only two numerical realizations of a disk galaxy within
clumpy dark matter halos, and it is always hazardous to extrapolate from such a
small number of cases. Our study does show, however, that it is possible at
least in some cases to maintain a stellar disk in spite of substructure. A
second caveat is that we have explored a model motivated by the present-day
structure of the Milky Way and by the $z=0$ substructure of a CDM halo. Models
that take into account the ongoing formation of the disk and a more realistic
treatment of the evolution of substructure are clearly desirable in order to
refine the conclusions presented here. Finally, for simplicity we have adopted a
non-singular isothermal model for the dark halo in the simulations reported
here, rather than the `cuspy' density profiles found in cosmological N-body
simulations (NFW). Although we do not expect this assumption to undermine the
conclusions reported here, we cannot rule out subtle effects that may arise from
the interaction between disk and halo in a cosmological context.

Finally, as discussed by Navarro \& Steinmetz (1997, 2000), it is quite
difficult to account simultaneously for the masses, luminosities, rotation
speeds, and angular momenta of galaxy disks in cosmogonies such as CDM, where
much of the mass of a virialized halo tends to be assembled through a sequence
of mergers. Until these issues are fully resolved it would be premature to
extend a clean bill of health to the CDM paradigm regarding the formation and
evolution of spiral galaxies like our own Milky Way.

 \acknowledgements

We would like to thank B.~Moore, F.~Governato, S.~Ghigna and G.~Lake, for
sharing with us an electronic version of the data showed in Figure 1. Lars
Hernquist kindly made available the software used to set up the equilibrium
disk/bulge/halo models. We thank the referee for a useful report that led to a
number of clarifications and improvements. The Natural Sciences \& Engineering
Research Council of Canada (NSERC) and the Canadian Foundation for Innovation
have supported this research through various grants to JFN.

\vfil\eject
\begin{figure}
\figurenum{3}
{\epsscale{1.} \plotone{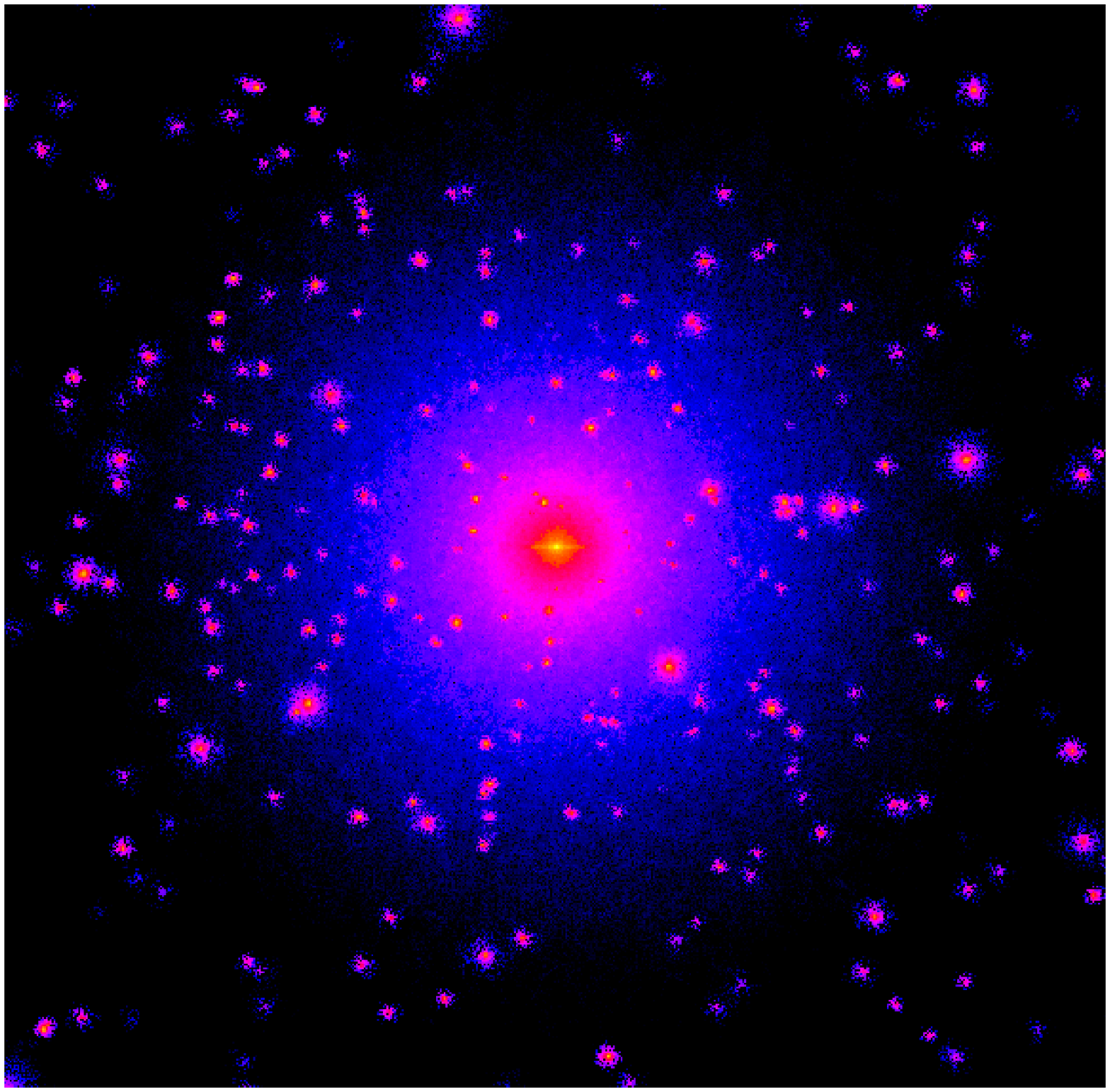}}
\caption {The disk/bulge/halo galaxy model including substructure halos. The
box is $600$ kpc on the side (about twice the virial radius) and shows the disk
and bulge of the Milky Way model at the center (in yellow) surrounded by a dark
halo populated with several hundred dark matter satellites. About half of the
subhalos are on prograde orbits. Note that because the halo extends well beyond
30 times the size of the luminous disk collisions between substructure halos and the stellar disk occur quite infrequently.}
{\label{fig3}} 
\end{figure}

\end{document}